\pdfoutput=1
\documentclass[12pt,preprint]{aastex}
\usepackage{graphicx}
\usepackage{natbib}
\usepackage{aas_macros}
\usepackage[section] {placeins}
\usepackage{subfigure}
\usepackage{float}
\usepackage{color}

\def\msun{{\rm\,M_\odot}}

\def\msun{{\rm\,M_\odot}}

\newcommand{\kms}{\, {\rm km\, s}^{-1}}

\newcommand{\mpc}{\, {\rm Mpc}}
\newcommand{\kpc}{\, {\rm kpc}}
\newcommand{\hmpc}{\, h^{-1} \mpc}

\newcommand{\hi}{\mbox{H\,{\scriptsize I}\ }}
\newcommand{\heii}{\mbox{He\,{\scriptsize II}\ }}

\def\h2{${\rm\,H_2}$}

\makeatletter 

\def\hi{\noindent \hangindent=2.5em}

\def\kpc{{\rm\,kpc}}

\def\mpc{{\rm\,Mpc}}

\def\kms{{\rm\,km/s}}
\def\msun{{\rm\,M_\odot}}

\def\vol#1  {{{#1}{\rm,}\ }}

\def\hi{{H\thinspace I}\ }

\def\heii{{He\thinspace II}\ }

\newcount\refno
\newcount\rfno
\def\eq{$^{\the\refno\ }$\advance\refno by 1}
\def\ad{\advance\rfno by 1}

\def\clock{\count0=\time \divide\count0 by 60
     \count1=\count0 \multiply\count1 by -60 \advance\count1 by \time
     \number\count0:\ifnum\count1<10{0\number\count1}\else\number\count1\fi}

\def\myputfigure#1#2#3#4#5%
{\hskip0.03\textwidth\vskip#5pt
\makebox[0pt]{\hskip#2in
\includegraphics[width=#3\textwidth]{#1}}\vskip#4pt\hfill}

\newcount\refno
\newcount\rfno
\def\eq{$^{\the\refno\ }$\advance\refno by 1}
\def\ad{\advance\rfno by 1}

\definecolor{burntorange}{rgb}{1,0.4,0.2}

\begin{document}

\title{Temporal Self-Organization in Galaxy Formation}

\author{Renyue Cen$^{1}$}

\footnotetext[1]{Princeton University Observatory, Princeton, NJ 08544;
 cen@astro.princeton.edu}

\begin{abstract}

We report on the discovery of a relation between the number of star formation (SF) peaks per unit time, $\nu_{\rm peak}$,
and the size of the temporal smoothing window function, $\Delta t$, used to define the peaks:
$\nu_{\rm peak}\propto\Delta t^{1-\phi}$ ($\phi\sim 1.618$).
This relation holds over the range of $\Delta t=10$ to $1000$Myr that can be reliably computed,
using a large sample of galaxies obtained from a state-of-the-art cosmological hydrodynamic simulation. 
This means that the temporal distribution of SF peaks in galaxies as a population
is fractal with a Hausdorff 
fractal dimension equal to $\phi-1$.
This finding reveals, for the first time, that the superficially chaotic process of galaxy formation 
is underlined by a temporal self-organization up to at least one gigayear.
It is tempting to suggest that, given the known existence of spatial fractals (such as the power-law two-point function of galaxies), 
there is a joint spatio-temporal self-organization in galaxy formation.
From an observational perspective, it will be urgent to devise diagnostics to probe SF histories of galaxies with good temporal resolution
to facilitate a test of this prediction.
If confirmed, it would provide unambiguous evidence for a new picture of galaxy formation that 
is interaction driven, cooperative and coherent in and between 
time and space. 
Unravelling its origin may hold the key to understanding galaxy formation.

\end{abstract}

\section{Introduction}\label{sec: intro}
Galaxy formation involves a large set of physical processes - cosmological expansion, gravity, hydrodynamics, atomic physics
and feedback from star formation, stellar evolution and black hole growth - and spans large dynamic ranges in time 
(at least $0.1$Myr to $10$Gyr) and space (at least $1$pc to $100$Mpc).
Some of the most interesting results on galaxy formation are thus obtained using large-scale simulations, providing fundamental insights 
on a variety of different aspects \citep[e.g.,][]{1988Frenk, 1994Cen, 
1998Gnedin, 1999Klypin, 1999Moore, 1999Cen, 2002Wechsler, 2002Abel, 2002Bromm, 2005Springel,
2005Keres, 2006Hopkins, 2006Croton, 2006Naab, 2007Bournaud, 2008Diemand, 2009Dekel, 2010Schaye}. 
The spatial distributions of galaxies have been extensively studied observationally, primarily at low redshift.
Among the most striking is the nature's ability to maintain a powerlaw galaxy-galay two-point correlation function 
over a significant range ($\sim 0.1-10h^{-1}$Mpc) \citep[e.g.,][]{1977Groth},
although there is evidence of a slight inflection at $\sim 1-2h^{-1}$Mpc in recent analysis \citep[e.g.,][]{2004Zehavi}.
This spatial regularity is not inherited from the linear power spectrum but must be a result of cooperation between nonlinear evolution and galaxy formation.
In self-gravitating systems, such as galaxies, the temporal and spatial structures may be related.
This may be seen by two examples.
First, for an isolated (non-dissipative) spherical system, the collapse time of each shell (assuming no shell crossings) 
is uniquely determined by the interior mass and specific energy of the shell that in turn is determined by the density structures.
Second, during the growth of a typical galaxy, in addition to direct acquisition of stars  via mergers and accretion (along with dark matter), 
significant spatial interactions may induce significant star formation activities hence leave temporal imprints in its star formation history.
Taking these indications together suggests that one should benefit by tackling the problem of galaxy formation 
combining the spatial and temporal information.
Here, as a step in that direction, we perform a novel analysis, utilizing the ab initio LAOZI adaptive mesh refinement cosmological hydrodynamic
simulation, to understand the statistical properties of star formation episodes in galaxies.

\section{Simulations}\label{sec: sims}

The reader is referred to \citet[][]{2014Cen} for detailed descriptions
of our simulations and the list of its empirical validations therein.
Briefly, a zoom-in region of comoving size of 
$21\times 24\times 20h^{-3}$Mpc$^3$
is embedded in a $120h^{-1}$Mpc periodic box 
and resolved to better than $114h^{-1}$pc (physical).
We use the following cosmological parameters that are consistent with
the WMAP7-normalized \citep[][]{2011Komatsu} $\Lambda$CDM model:
$\Omega_M=0.28$, $\Omega_b=0.046$, $\Omega_{\Lambda}=0.72$, $\sigma_8=0.82$,
$H_0=100 h \,{\rm km\, s}^{-1} {\rm Mpc}^{-1} = 70 \,{\rm km\, s}^{-1} {\rm Mpc}^{-1}$ and $n=0.96$.
Equations governing motions of dark matter, gas and stars, and thermodynamic state of gas are followed forward in time from redshift $100$ to $0.62$,
using the adaptive mesh refinement cosmological hydrodynamic code Enzo \citep[][]{2013Enzo},
which includes all important microphysics and major feedback processes that are well measured.
Stellar particles (equivalent to coeval stellar cluster of mass $\sim 10^5\msun$) 
are created from gas clouds meeting certain physical conditions over time, 
based on the empirical Kennicutt-Schmidt law \citep[][]{1998Kennicutt}.
Stellar particles at any time may be grouped together spatially  
using the HOP algorithm \citep[][]{1998Eisenstein} to create galaxy catalogs, which are tested to be robust
and insensitive to specific choices of concerned parameters within reasonable ranges.
For each galaxy we have its exact star formation history, given its member stellar particles formation times.
A total of (2090, 965, 296, 94, 32, 10) galaxies are found with stellar masses greater than 
($10^{9.5}, 10^{10}, 10^{10.5}, 10^{11}, 10^{11.5}, 10^{12})\msun$ at $z=0.62$.

For each galaxy we create an uniform time grid of star formation rate at a time resolution of $3$Myr
from redshift $20$ to $0.62$, which we call the ``unsmoothed" SF history, denoted as $S(t)$.
We then smooth $S(t)$ using a square window of full width equal to $t_{s}$ to create a locally-averaged version, denoted as $\bar S(t)$, 
which is defined to be $\bar S(t)\equiv {1\over t_s} \int_{t-t_s/2}^{t+t_s/2} S(t^\prime) dt^\prime$. 
Another variable is then defined from $\bar S(t)$: $\delta (t) \equiv S(t)-\bar S(t)$.
We smooth $\delta (t)$ with a gaussian window of radius $t_{\rm g}$ to yield $\bar{\delta}(t)$. 
We obtain finally $S_s(t) \equiv \bar S(t) + \bar{\delta}(t)$. 
We identify SF peaks in $S_s(t)$ as follows.
Each SF peak is defined as a contiguous region between two consecutive
local minima in $S_s(t)$, say, at time $t_1$ and $t_2$.  
We sum up $S(t)$ in the same temporal region [$t_1$, $t_2$] to get the total stellar mass for the peak.
For each galaxy, we catalog and rank order a complete list of peaks each containing the following information:
the total stellar mass, the point in time of maximum SFR and the rank.
The number of top SF peaks that make up 50\% and 90\% of total amount of stellar mass of a galaxy at $z=0.62$
is denoted, $n_{50}$ and $n_{90}$, respectively.
We note that the main purpose of smoothing 
$\delta (t)$ with the gaussian window is to make the automated peak identification method umambiguous. 
Thus, it is $t_s$ that serves as a time ``ruler".
We use $t_g=t_s/2$ and find the slope of the scaling relation found does not depend on $t_s/t_g$ within the concerned accuracies.

\begin{figure}[!h]
\centering
\vskip -0.3cm
\resizebox{5.0in}{!}{\includegraphics[angle=0]{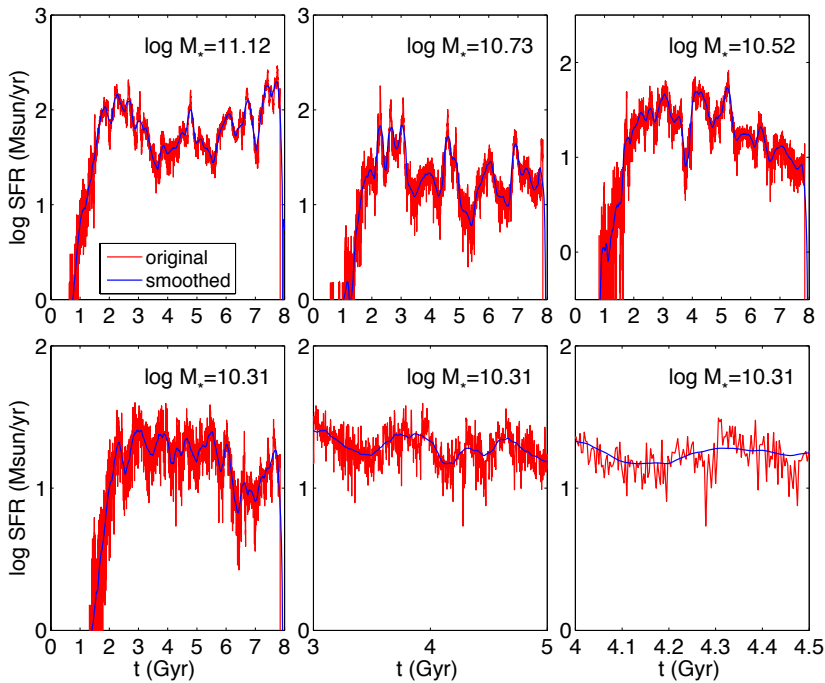}}   
\vskip -1.0cm
\caption{
shows the star formation histories for four galaxies (the top row plus the bottom-left panel)
selected semi-randomly covering mass range of interest at $z=0.62$.
The time starts at the big bang as zero. 
The red curves are for unsmoothed SF histories $S(t)$.
The blue curves are for the corresponding smoothed SF histories $S_s(t)$,
with $t_s=200~$Myr. 
In each panel, the galaxy stellar mass at $z=0.62$ is indicated at the top.
The bottom-middle and -right panels are zoom-in views of the same galaxy shown in the bottom-left panel.
}
\label{fig:sfr}
\end{figure}

\section{Results}\label{sec: results}

\begin{figure}[!b]
\centering
\vskip -0.5cm
\resizebox{5.0in}{!}{\includegraphics[angle=0]{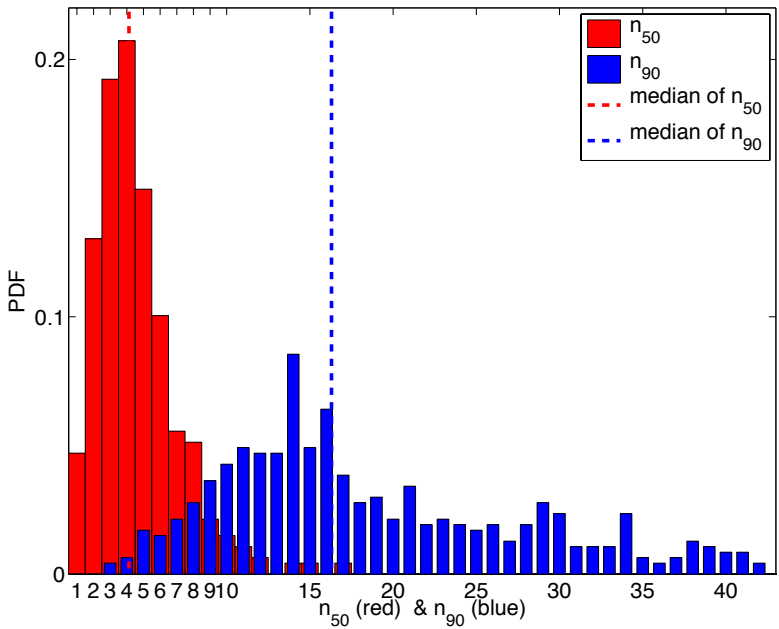}}   
\vskip -1.0cm
\caption{
shows the probability distribution function (PDF)
of the number of top SF peaks contributing to 50\% ($n_{50}$) and 90\% ($n_{90}$), 
respectively, of total stellar mass at $z=0.62$ 
for all galaxies more massive  than $10^{10}\msun$.
The vertical red and and blue dashed lines indicate the median of the respective historgrams.
The peaks are identified with $t_s=200$Myr.
}
\label{fig:perc50}
\end{figure}

We start by showing the star formation histories for four galaxies  
in Figure~\ref{fig:sfr}.
We see that our adaptive smoothing scheme appropriately retains major SF peaks 
but smooths out high-frequency peaks on scales smaller than the ruler size $t_s$, exactly serving the purpose.
We also see that there are temporal structures from $\sim 1$Myr to $\sim 1$Gyr.
Although it is difficult to quantify visually the nature of the temporal structures,
there is a hint that a significant SF peak is often sandwiched by periods of 
diminished SF activities or less significant SF peaks.
It is evident that the histories of individual galaxies vary substantially with respect to 
both the trend on long time scales and fluctuations on short time scales.
Anectodal evidence that is consistent with the global evolution of SFR density \citep[][]{2006HopkinsA}
is that, for the galaxy population as a whole, the majority of galaxies are on a downward trend of SFR with increasing time (decreasing redshift)
from $t\sim 2-3$Gyr (corresponding to $z=2$ to $3$).
It is seen that SF in galaxies is usually not monolithic. 
A typical galaxy is found to have a polylithic temporal structure of star formation,
consisting of a series of quasi-monoliths occurring in time in an apparently chaotic fashion. 
Not only is there no evidence that a typical galaxy forms most of its stars in a single burst,
but also the SF history over any scale does not display a form that may be represented 
by any simple analytic functions (such as an exponential).
A qualitatively similar appearance of oscillatory star formation rates are seen in \citet[][]{2013Hopkins},
although detailed quantitative comparisons are not available at this time.
One take-away message is this: galaxy formation is a chaotic process and 
conclusions about the galaxy population as a whole based on an unrepresentative sample of galaxies should be taken cautiously.
Another is that the often adopted simple temporal profiles for star formation (such as exponential decay or delta function)
in interpreting observational results should be reconsidered.

\begin{figure}[!h]
\centering
\vskip -0.3cm
\resizebox{5.0in}{!}{\includegraphics[angle=0]{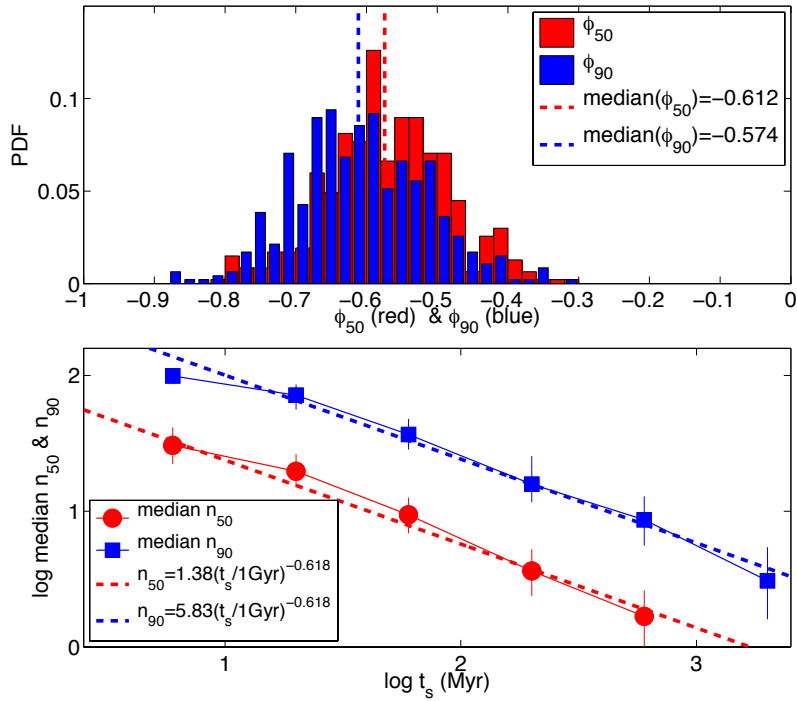}}   
\vskip -0.8cm
\caption{
Top panel shows the PDF of 
${\phi_{50}}$ (red histogram) and ${\phi_{90}}$ (blue historgram) 
in the fit $n_{50}\propto t_s^{\phi_{50}}$ and $n_{90}\propto t_s^{\phi_{90}}$
for all galaxies with stellar masses greater than $10^{10}\msun$ at $z=0.62$.
The vertical red and and blue dashed lines indicate the median of the red and blue historgrams, respectively.
Bottom panel shows the median of $n_{50}$ (red dots) and $n_{90}$ (blue squares), respectively,
for all galaxies with stellar masses greater than $10^{10}\msun$ at $z=0.62$,
as a function of temporal smoothing window $t_s$.
The vertical errorbars indicate the 25\%-75\% range.
The red and and blue dashed lines indicate fits with a slope $-0.618$.
}
\label{fig:slope}
\end{figure}

We now turn to quantitative results.
Figure~\ref{fig:perc50} shows the PDFs of $n_{50}$ and $n_{90}$ with $t_s=200$Myr.
We see that the number of peaking containing 50\% of stellar mass ($n_{50}$) falls in the range of $\sim 1-10$ peaks, 
whereas the number of peaks containing 90\% of stellar mass ($n_{90}$) displays a much broader range of $\sim 5-40$.
We note that, had we restricted the galaxy stellar mass range to 
$10^{10-11}$ or $10^{11-12}\msun$,
the results do not change significantly.
It is clear that there are large variations from galaxy to galaxy with respect to individual SF histories,
as was already hinted in in Figure~\ref{fig:sfr}.
Behind this chaos, however, collectively, an order is found, as will be shown in Figure~\ref{fig:slope}.

\begin{figure}[!h]
\centering
\vskip -0.3cm
\resizebox{5.0in}{!}{\includegraphics[angle=0]{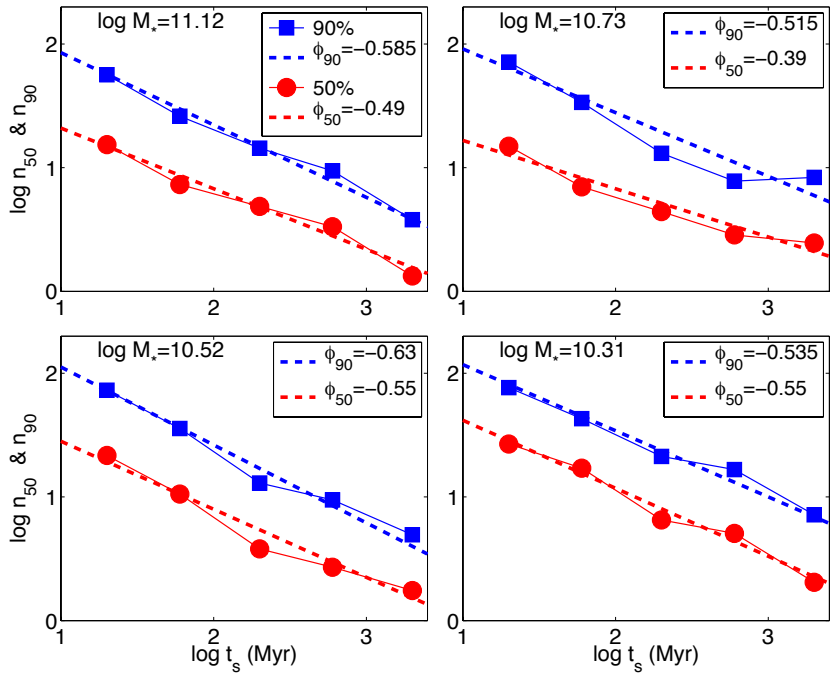}}   
\vskip -0.8cm
\caption{
shows $n_{50}$ (red dots) and $n_{90}$ (blue squares) as a function of 
temporal smoothing window $t_s$ for the four galaxies shown in Figure~\ref{fig:sfr}.
Linear fits to the $\log t_s$ - $\log n_{50}$ 
and $\log t_s$ - $\log n_{90}$ are shown as dashed lines with the respective colors.
}
\label{fig:Ntindiv}
\end{figure}

Figure~\ref{fig:Ntindiv} shows $n_{50}$ (red dots) and $n_{90}$ (blue squares) as a function of 
temporal smoothing window $t_s$ for the four galaxies shown in Figure~\ref{fig:sfr}.
We see that powerlaw fits - $n_{50}\propto t_s^{\phi_{50}}$ and $n_{90}\propto t_s^{\phi_{90}}$ -
provide reasonable approximations.
Collecting all galaxies with stellar masses greater than $10^{10}\msun$ at $z=0.62$
the results are shown in Figure~\ref{fig:slope}. 
The top panel of Figure~\ref{fig:slope} shows
the PDF of ${\phi_{50}}$ (red histogram) and ${\phi_{90}}$ (blue historgram). 
We see that there are substantial variations among galaxies, which is expected.
The most significant point is that a typical galaxy has 
$\phi_{50}$ and $\phi_{90}$ around $-0.6$.
In other words, the galaxy population, collectively taken as a whole, displays significant orderliness.
This point is re-enforced in the bottom panel of Figure~\ref{fig:slope}, which is similar to  
Figure~\ref{fig:Ntindiv}.
But here, instead of showing powerlaw fits for individual galaxies,
we compute the median of  $n_{50}$ (red dots) and $n_{90}$ (blue squares) for all galaxies first as a function of $t_s$ and then show the fits
to the medians.
It is intriguing that a slope about $-0.618$ ($=1-\phi$) provides a quite good fit,
where $\phi=1.618$ is often called the golden ratio.

\section{Discussion and Conclusions}\label{sec: conclusions}

This paper is the third in the series ``On the Origin of the Hubble Sequence".
Utilizing {\it ab initio}
{\color{red}\bf L}arge-scale {\color{red}\bf A}daptive-mesh-refinement
{\color{red}\bf O}mniscient {\color{red}\bf Z}oom-{\color{red}\bf I}n cosmological hydrodynamic simulations
({\color{red}\bf LAOZI Simulations}) of the standard cold dark matter model,
we undertake a unique study of the statistical properties of star formation episodes in galaxies.
We find  a relation between the number of star formation (SF) peaks per unit time, $\nu_{\rm peak}$,
and the size of the temporal smoothing window function, $\Delta t$, used to define the peaks:
$\nu_{\rm peak}\propto\Delta t^{1-\phi}$ ($\phi\sim 1.618$),
valid over the range of $\Delta t=0.01-1$Gyr.
It is expected that the findings do not significantly depend on precise cosmological parameters, since
the responsible processes are mostly in the nonlinear regime,
although it remains to be seen if the relation extends to below 10Myr, where non-gravitational processes, including feedback processes,
may introduce time scales of their own.
The implication is profound: galaxy formation is temporally fractal and displays a self-organization up to at least one gigayear,
with a \citet[][]{1919Hausdorff} dimension equal to $\phi-1$. 

We attribute this temporal self-organization, tentatively, to interactions between galaxies that presumably trigger star formation peaks
and are organized temporally in a way that is yet to be quantitatively understood.
Qualitatively, the found results may be explained as follows.
One could envision that galaxies are normally (at least at high redshift) embedded in a gas reservoir,
which is the potential fuel of star formation.
When there is a trigger, some of this gas is driven inward to fuel star formation.
The triggers are likely due to significant interactions between galaxies,
such as major and minor mergers or close fly-bys of significant galaxies, 
or some torquing events,
or some hydrodynamic events.
The triggers may be democratically distributed temporally in the sense that
at a given time baseline a large trigger is not usually preceded or followed by another large trigger,
but rather by small triggers. 
One might even argue that in some rare cases, even if a large trigger does follow a preceding large one,
a significant ``drawdown" of gas by the preceding SF peak may cause the second SF peak
to be less powerful that it otherwise would.
Such compensated behavior could give rise to the temporal structures seen.
Were the triggers distributed randomly, then ${\phi}$ would be $2$.
Should the triggers be completely correlated (i.e., a delta function in time), 
then ${\phi}$ would be $1$.

Since the triggering of SF peaks by galaxy interactions implies spatial correlations of galaxies,
and given that galaxies are known to exhibit spatial fractals, such as the power-law galaxy two-point correlation function \citep[e.g.,][]{1980Peebles},
our results are strongly indicative that galaxy formation may be governed by a fundamental joint spatio-temporal self-organization.
Understanding the origin of this self-organization may hold a key to understanding galaxy formation.

Observational diagnostics to probe SF histories of galaxies with competitively good temporal resolution
from a few Myr to Gyr, especially those that are applicable to a sufficient sample of galaxies,
are highly wanted, in order to test the predictions made here.
In addition, with the development of this new line of inquiry, more accurate observational 
characterizations of galaxy clustering at high redshift
at the peak of star formation will be useful.

In spite of the apparent coincidence, it would be premature to emphatically relate $\phi$ 
to the golden ratio. 
Nonetheless, the ubiquitous manifestations of the golden ratio in nature 
suggest that further investigations with higher statistical accuracies may be warranted.
Could the galaxy formation be golden after all?

\vskip 1cm
I would like to thank Claire Lackner for providing the SQL based merger tree construction software.
The analysis program yt \citep[][]{2011Turk} is used to perform some of the analysis.
Computing resources were in part provided by the NASA High-
End Computing (HEC) Program through the NASA Advanced
Supercomputing (NAS) Division at Ames Research Center.
This work is supported in part by grant NASA NNX11AI23G.


\begin{thebibliography}{30}
\expandafter\ifx\csname natexlab\endcsname\relax\def\natexlab#1{#1}\fi

\bibitem[{{Abel} {et~al.}(2002){Abel}, {Bryan}, \& {Norman}}]{2002Abel}
{Abel}, T., {Bryan}, G.~L., \& {Norman}, M.~L. 2002, Science, 295, 93

\bibitem[{{Bournaud} {et~al.}(2007){Bournaud}, {Elmegreen}, \&
  {Elmegreen}}]{2007Bournaud}
{Bournaud}, F., {Elmegreen}, B.~G., \& {Elmegreen}, D.~M. 2007, \apj, 670, 237

\bibitem[{{Bromm} {et~al.}(2002){Bromm}, {Coppi}, \& {Larson}}]{2002Bromm}
{Bromm}, V., {Coppi}, P.~S., \& {Larson}, R.~B. 2002, \apj, 564, 23

\bibitem[{{Cen}(2014)}]{2014Cen}
{Cen}, R. 2014, \apj, 781, 38

\bibitem[{{Cen} {et~al.}(1994){Cen}, {Miralda-Escude}, {Ostriker}, \&
  {Rauch}}]{1994Cen}
{Cen}, R., {Miralda-Escude}, J., {Ostriker}, J.~P., \& {Rauch}, M. 1994, \apjl,
  437, L9

\bibitem[{{Cen} \& {Ostriker}(1999)}]{1999Cen}
{Cen}, R., \& {Ostriker}, J.~P. 1999, \apj, 514, 1

\bibitem[{{Croton} {et~al.}(2006){Croton}, {Springel}, {White}, {De Lucia},
  {Frenk}, {Gao}, {Jenkins}, {Kauffmann}, {Navarro}, \& {Yoshida}}]{2006Croton}
{Croton}, D.~J., {Springel}, V., {White}, S.~D.~M., {De Lucia}, G., {Frenk},
  C.~S., {Gao}, L., {Jenkins}, A., {Kauffmann}, G., {Navarro}, J.~F., \&
  {Yoshida}, N. 2006, \mnras, 365, 11

\bibitem[{{Dekel} {et~al.}(2009){Dekel}, {Birnboim}, {Engel}, {Freundlich},
  {Goerdt}, {Mumcuoglu}, {Neistein}, {Pichon}, {Teyssier}, \&
  {Zinger}}]{2009Dekel}
{Dekel}, A., {Birnboim}, Y., {Engel}, G., {Freundlich}, J., {Goerdt}, T.,
  {Mumcuoglu}, M., {Neistein}, E., {Pichon}, C., {Teyssier}, R., \& {Zinger},
  E. 2009, \nat, 457, 451

\bibitem[{{Diemand} {et~al.}(2008){Diemand}, {Kuhlen}, {Madau}, {Zemp},
  {Moore}, {Potter}, \& {Stadel}}]{2008Diemand}
{Diemand}, J., {Kuhlen}, M., {Madau}, P., {Zemp}, M., {Moore}, B., {Potter},
  D., \& {Stadel}, J. 2008, \nat, 454, 735

\bibitem[{Eisenstein \& Hut(1998)}]{1998Eisenstein}
Eisenstein, D.~J., \& Hut, P. 1998, ApJ, 498, 137

\bibitem[{{Frenk} {et~al.}(1988){Frenk}, {White}, {Davis}, \&
  {Efstathiou}}]{1988Frenk}
{Frenk}, C.~S., {White}, S.~D.~M., {Davis}, M., \& {Efstathiou}, G. 1988, \apj,
  327, 507

\bibitem[{{Gnedin}(1998)}]{1998Gnedin}
{Gnedin}, N.~Y. 1998, \mnras, 294, 407

\bibitem[{{Groth} \& {Peebles}(1977)}]{1977Groth}
{Groth}, E.~J., \& {Peebles}, P.~J.~E. 1977, \apj, 217, 385

\bibitem[{Hausdorff(1919)}]{1919Hausdorff}
Hausdorff, F. 1919, Mathematische Annalen, 79, 157

\bibitem[{{Hopkins} \& {Beacom}(2006)}]{2006HopkinsA}
{Hopkins}, A.~M., \& {Beacom}, J.~F. 2006, ApJ, 651, 142

\bibitem[{{Hopkins} {et~al.}(2006){Hopkins}, {Hernquist}, {Cox}, {Di Matteo},
  {Robertson}, \& {Springel}}]{2006Hopkins}
{Hopkins}, P.~F., {Hernquist}, L., {Cox}, T.~J., {Di Matteo}, T., {Robertson},
  B., \& {Springel}, V. 2006, \apjs, 163, 1

\bibitem[{{Hopkins} {et~al.}(2013){Hopkins}, {Keres}, {Onorbe},
  {Faucher-Giguere}, {Quataert}, {Murray}, \& {Bullock}}]{2013Hopkins}
{Hopkins}, P.~F., {Keres}, D., {Onorbe}, J., {Faucher-Giguere}, C.-A.,
  {Quataert}, E., {Murray}, N., \& {Bullock}, J.~S. 2013, ArXiv e-prints

\bibitem[{{Kennicutt}(1998)}]{1998Kennicutt}
{Kennicutt}, Jr., R.~C. 1998, \apj, 498, 541

\bibitem[{{Kere{\v s}} {et~al.}(2005){Kere{\v s}}, {Katz}, {Weinberg}, \&
  {Dav{\'e}}}]{2005Keres}
{Kere{\v s}}, D., {Katz}, N., {Weinberg}, D.~H., \& {Dav{\'e}}, R. 2005,
  \mnras, 363, 2

\bibitem[{{Klypin} {et~al.}(1999){Klypin}, {Kravtsov}, {Valenzuela}, \&
  {Prada}}]{1999Klypin}
{Klypin}, A., {Kravtsov}, A.~V., {Valenzuela}, O., \& {Prada}, F. 1999, \apj,
  522, 82

\bibitem[{{Komatsu} {et~al.}(2011){Komatsu}, {Smith}, {Dunkley}, {Bennett},
  {Gold}, {Hinshaw}, {Jarosik}, {Larson}, {Nolta}, {Page}, {Spergel},
  {Halpern}, {Hill}, {Kogut}, {Limon}, {Meyer}, {Odegard}, {Tucker}, {Weiland},
  {Wollack}, \& {Wright}}]{2011Komatsu}
{Komatsu}, E., {Smith}, K.~M., {Dunkley}, J., {Bennett}, C.~L., {Gold}, B.,
  {Hinshaw}, G., {Jarosik}, N., {Larson}, D., {Nolta}, M.~R., {Page}, L.,
  {Spergel}, D.~N., {Halpern}, M., {Hill}, R.~S., {Kogut}, A., {Limon}, M.,
  {Meyer}, S.~S., {Odegard}, N., {Tucker}, G.~S., {Weiland}, J.~L., {Wollack},
  E., \& {Wright}, E.~L. 2011, \apjs, 192, 18

\bibitem[{{Moore} {et~al.}(1999){Moore}, {Ghigna}, {Governato}, {Lake},
  {Quinn}, {Stadel}, \& {Tozzi}}]{1999Moore}
{Moore}, B., {Ghigna}, S., {Governato}, F., {Lake}, G., {Quinn}, T., {Stadel},
  J., \& {Tozzi}, P. 1999, \apjl, 524, L19

\bibitem[{{Naab} {et~al.}(2006){Naab}, {Khochfar}, \& {Burkert}}]{2006Naab}
{Naab}, T., {Khochfar}, S., \& {Burkert}, A. 2006, \apjl, 636, L81

\bibitem[{{Peebles}(1980)}]{1980Peebles}
{Peebles}, P.~J.~E. 1980, {The large-scale structure of the universe} (Research
  supported by the National Science Foundation.~Princeton, N.J., Princeton
  University Press, 1980.~435 p.)

\bibitem[{{Schaye} {et~al.}(2010){Schaye}, {Dalla Vecchia}, {Booth}, {Wiersma},
  {Theuns}, {Haas}, {Bertone}, {Duffy}, {McCarthy}, \& {van de
  Voort}}]{2010Schaye}
{Schaye}, J., {Dalla Vecchia}, C., {Booth}, C.~M., {Wiersma}, R.~P.~C.,
  {Theuns}, T., {Haas}, M.~R., {Bertone}, S., {Duffy}, A.~R., {McCarthy},
  I.~G., \& {van de Voort}, F. 2010, \mnras, 402, 1536

\bibitem[{{Springel} {et~al.}(2005){Springel}, {White}, {Jenkins}, {Frenk},
  {Yoshida}, {Gao}, {Navarro}, {Thacker}, {Croton}, {Helly}, {Peacock}, {Cole},
  {Thomas}, {Couchman}, {Evrard}, {Colberg}, \& {Pearce}}]{2005Springel}
{Springel}, V., {White}, S.~D.~M., {Jenkins}, A., {Frenk}, C.~S., {Yoshida},
  N., {Gao}, L., {Navarro}, J., {Thacker}, R., {Croton}, D., {Helly}, J.,
  {Peacock}, J.~A., {Cole}, S., {Thomas}, P., {Couchman}, H., {Evrard}, A.,
  {Colberg}, J., \& {Pearce}, F. 2005, \nat, 435, 629

\bibitem[{{The Enzo Collaboration} {et~al.}(2013){The Enzo Collaboration},
  {Bryan}, {Norman}, {O'Shea}, {Abel}, {Wise}, {Turk}, {Reynolds}, {Collins},
  {Wang}, {Skillman}, {Smith}, {Harkness}, {Bordner}, {Kim}, {Kuhlen}, {Xu},
  {Goldbaum}, {Hummels}, {Kritsuk}, {Tasker}, {Skory}, {Simpson}, {Hahn},
  {Oishi}, {So}, {Zhao}, {Cen}, \& {Li}}]{2013Enzo}
{The Enzo Collaboration}, {Bryan}, G.~L., {Norman}, M.~L., {O'Shea}, B.~W.,
  {Abel}, T., {Wise}, J.~H., {Turk}, M.~J., {Reynolds}, D.~R., {Collins},
  D.~C., {Wang}, P., {Skillman}, S.~W., {Smith}, B., {Harkness}, R.~P.,
  {Bordner}, J., {Kim}, J.-h., {Kuhlen}, M., {Xu}, H., {Goldbaum}, N.,
  {Hummels}, C., {Kritsuk}, A.~G., {Tasker}, E., {Skory}, S., {Simpson}, C.~M.,
  {Hahn}, O., {Oishi}, J.~S., {So}, G.~C., {Zhao}, F., {Cen}, R., \& {Li}, Y.
  2013, ArXiv e-prints

\bibitem[{{Turk} {et~al.}(2011){Turk}, {Smith}, {Oishi}, {Skory}, {Skillman},
  {Abel}, \& {Norman}}]{2011Turk}
{Turk}, M.~J., {Smith}, B.~D., {Oishi}, J.~S., {Skory}, S., {Skillman}, S.~W.,
  {Abel}, T., \& {Norman}, M.~L. 2011, \apjs, 192, 9

\bibitem[{{Wechsler} {et~al.}(2002){Wechsler}, {Bullock}, {Primack},
  {Kravtsov}, \& {Dekel}}]{2002Wechsler}
{Wechsler}, R.~H., {Bullock}, J.~S., {Primack}, J.~R., {Kravtsov}, A.~V., \&
  {Dekel}, A. 2002, \apj, 568, 52

\bibitem[{{Zehavi} {et~al.}(2004){Zehavi}, {Weinberg}, {Zheng}, {Berlind},
  {Frieman}, {Scoccimarro}, {Sheth}, {Blanton}, {Tegmark}, {Mo}, {Bahcall},
  {Brinkmann}, {Burles}, {Csabai}, {Fukugita}, {Gunn}, {Lamb}, {Loveday},
  {Lupton}, {Meiksin}, {Munn}, {Nichol}, {Schlegel}, {Schneider}, {SubbaRao},
  {Szalay}, {Uomoto}, {York}, \& {SDSS Collaboration}}]{2004Zehavi}
{Zehavi}, I., {Weinberg}, D.~H., {Zheng}, Z., {Berlind}, A.~A., {Frieman},
  J.~A., {Scoccimarro}, R., {Sheth}, R.~K., {Blanton}, M.~R., {Tegmark}, M.,
  {Mo}, H.~J., {Bahcall}, N.~A., {Brinkmann}, J., {Burles}, S., {Csabai}, I.,
  {Fukugita}, M., {Gunn}, J.~E., {Lamb}, D.~Q., {Loveday}, J., {Lupton}, R.~H.,
  {Meiksin}, A., {Munn}, J.~A., {Nichol}, R.~C., {Schlegel}, D., {Schneider},
  D.~P., {SubbaRao}, M., {Szalay}, A.~S., {Uomoto}, A., {York}, D.~G., \& {SDSS
  Collaboration}. 2004, \apj, 608, 16

\end{thebibliography}

\end{document}